\documentclass[12pt,eqsecnum,floats,aps,amsmath,amssymb,nofootinbib,prd]{article}

\usepackage{authblk}
\usepackage{setspace}
\usepackage{comment}
\usepackage{setspace}
\usepackage{amsmath,amssymb,amsfonts,amsthm,mathrsfs}
\usepackage{graphicx}
\usepackage{enumerate} 
\usepackage[pdftex]{hyperref}

\def\be{\begin{equation}}
\def\ee{\end{equation}}
\def\ba{\begin{eqnarray}}
\def\ea{\end{eqnarray}}
\def\bi{\begin{itemize}}
\def\ei{\end{itemize}}

\def\bra{\langle}
\def\ket{\rangle}

\def\O{\Omega}
\def\xh{\hat{x}}
\def\A{\mathcal{A}}

\def\I{\mathcal{I}}
\def\e{\varepsilon}
\def\qh{\hat{q}}

\def\wb{\bar{w}}
\def\zb{\bar{z}}
\def\w{\omega}
\def\t{\tau}
\def\H{\mathcal{H}}
\def\Qs{Q^{\rm soft}}
\def\Qh{Q^{\rm hard}}
\def\F{\mathcal{F}}
\def\J{\mathcal{J}}
\def\lamt{\tilde{\lambda}}
\def\laminf{\lambda_\H}
\def\G{\mathcal{G}}
\title{Asymptotic symmetries of QED and Weinberg's soft photon theorem}
\author[a]{Miguel Campiglia}
\author[b]{Alok Laddha}
\affil[a]{Universidad de la Rep\'ublica, Montevideo, Uruguay}
\affil[b]{Chennai Mathematical Institute, Chennai, India}

\begin{document}
\maketitle

\thispagestyle{empty}

\let\oldthefootnote\thefootnote
\renewcommand{\thefootnote}{\fnsymbol{footnote}}
\footnotetext{Email: campi@fisica.edu.uy, aladdha@cmi.ac.in}
\let\thefootnote\oldthefootnote

\begin{abstract}

Various equivalences between  so-called soft theorems which constrain  scattering amplitudes 
and  Ward identities related to  asymptotic symmetries have recently  been established in gauge theories and gravity. So far these equivalences have been restricted to the case of massless matter fields, the reason being that the asymptotic symmetries are defined at null infinity. The restriction is however unnatural from the perspective of  soft theorems which are insensitive to the masses of the external particles. 

In this work we  remove the aforementioned restriction in the context of scalar QED. Inspired by the radiative phase space description of massless fields at null infinity,  we introduce a manifold description of  time-like infinity on which the  asymptotic phase space for massive fields can be defined.  The ``angle dependent'' large gauge transformations are  shown to have a well defined action on this phase space, and the resulting Ward identities are found to be equivalent to Weinberg's soft photon theorem.

\end{abstract}
\maketitle

\section{Introduction}

Recently a new perspective on asymptotic symmetries and infrared effects in quantum field theories on (asymptotically) flat space times has emerged \cite{strom1,stromBMS1,stromBMS2,stromBMS3,stromqed}. The presence of massless particles (responsible for the infrared effects) implies the existence of a radiative phase space associated to such a theory which is localized at null infinity \cite{aaprl,aaradiative,AS,aaoxf,aa}. The asymptotic symmetries are ``large gauge transformations'' which do not die down at infinity and give rise to charges defined on this phase space. The Ward identities associated to such charges turn out to be related to the so-called soft theorems which are infrared constraints on scattering amplitudes. 

In the case of  Quantum Electrodynamics (QED), it was shown in a series of seminal works by Strominger and He et. al. \cite{strom1,stromqed} that the soft photons of the theory are goldstone bosons associated to a spontaneous breaking of asymptotic symmetries. The connection between asymptotic symmetries and soft photons was firmly established by proving an equivalence between Weinberg's soft photon theorem \cite{weinberg} and the Ward identities associated to the (spontaneously broken) large gauge transformations \cite{stromqed}.\\
This work was further generalized in a number of different directions, for instance by clarifying the connection between the symmetry group of semiclassical gravity in asymptotically flat spacetimes (known as the BMS group \cite{AS,aa}) and various soft graviton theorems which arise in perturbative quantum gravity \cite{stromBMS1,stromBMS2,stromBMS3,cl1,cl2}.\\
Remarkable as this progress has been, it has so far remained restricted to situations where all the particles involved in the scattering are massless. 
However, as Weinberg's soft theorems are applicable (and in-fact originally derived for!) theories where the photons or gravitons couple to  massive particles, it is a natural question to ask if the connection between asymptotic symmetries and soft theorem remains valid in that case. In this paper, we show that in the case of scalar QED where the charged scalar particles are massive,  this is indeed the case. \emph{That is, the existence of an infinite dimensional Abelian group of asymptotic symmetries of the (formal) S-matrix of QED is equivalent to  Weinberg's soft photon theorem.}\\
In this regard we should note that building upon the seminal work of Frohlich et. al. \cite{frohlich}, Balachandran et. al. have already made significant progress in analyzing the relationship between the group of large gauge transformations and infrared effects in QED \cite{balsachin0,balsachin}. This group of large gauge transformations is known as the Sky group and the corresponding charges are defined as integral over Cauchy slices in spacetime which intersect  spatial-infinity at $i^{0}$. Despite some subtle technical differences between the Sky group and the group of asymptotic symmetries as defined in the present work, we believe that they are different manifestations of the same group of large gauge transformations. However clarifying this relationship is a question we leave for the future.\\
We consider  scalar QED\footnote{Henceforth by scalar QED we would mean the theory where charged scalar particles are taken to be massive.} in  flat space-time. Weinberg's soft theorem can be written as a relation between scattering amplitudes where the massive scattering states are defined at future or past time-like infinity and massless (photon) states defined at null infinity. We show that this theorem is equivalent to Ward identities associated to the infinite dimensional group of  ``angle dependent" large gauge transformations of QED.\\
The paper is divided into the following parts. In section two, we define what we call the asymptotic phase space of a  massive complex scalar field at time-like infinity. 
This phase space is analogue to the radiative phase space of a massless scalar field as derived by Ashtekar and Strubel \cite{AS}. One subtlety  is that time-like infinity is usually described by  a point and hence  one cannot associate any phase space to it. 
We thus consider a description of time-like infinity as a unit space-like hyperboloid ${\cal H}^{\pm}$. This description is analogous to that of space-like infinity as defined by Ashtekar and Romano \cite{AR}. We then define a phase space of massive fields which are defined on ${\cal H}^{\pm}$. Quantization of this phase space leads to  free (charged) scattering states of the massive scalar field.\\ 
In section three we describe the  asymptotic phase space of the interacting Maxwell + scalar field system.
We work in Lorenz gauge, as the derivation of soft theorem is most succinctly done in this gauge. We show that even in Lorenz gauge the radiative phase space of the Maxwell field remains the same as the one obtained by working in radial gauge \cite{stromqed}. Thus the total ``asymptotic phase space" of the system is 
as a direct product of the radiative phase space of Maxwell field, defined at ${\cal I}^{\pm}$, and the asymptotic phase space of massive scalar field which is defined at the three manifold ${\cal H}^\pm$ that represents  time-like infinity.\\
In section four we derive the generators of large gauge transformations in the total asymptotic phase space of scalar QED. We show that this generator is a sum of two parts precisely as in the massless case \cite{AS,stromqed}. There is a term proportional to the time component of the matter current which is an integral over a three dimensional space-like hyperboloid and there is a term which is linear in the soft photon field that generates large gauge transformations of the Maxwell field. 
\\
In section five we write down the Ward identities associated to such large gauge transformations and show the equivalence between these identities and Weinberg's soft photon theorem.  In section six we summarize the results and make some comments on applying the present ideas to perturbative gravity.

\section{Asymptotics of a free massive scalar field}
Consider a free complex scalar field in Minkowski space:
\be
\varphi(x)  =  \int \frac{d^3 \vec{p}}{ (2 \pi)^3 2 E_p} ( b(\vec{p}) e^{i p \cdot x} + d^*(\vec{p}) e^{- i p \cdot x}), \label{freephi}
\ee
where $E_p=\sqrt{|\vec{p}|^2+m^2}$, $p \cdot x = - E_p t + \vec{p} \cdot \vec{x}$. In quantum theory $b(\vec{p})$ and $d(\vec{p})$  become the annihilation operators for the particles and antiparticles respectively.

We would like to obtain the `free data' $b(\vec{p})$ and $d(\vec{p})$ from a late time asymptotics of the field $\varphi(x)$.\footnote{For concreteness we focus on future asymptotics, analogous discussion can be made for asymptotics in the past.} In the $m=0$ case this is accomplished by considering a large time limit along null geodesics, which results in an identification of $b(\vec{p})$ and $d(\vec{p})$ with radiative phase space  data  of the field at future null infinity $\I^+$ (see for instance \cite{frolov,dp}).

For the present $m>0$ case, the appropriate late time asymptotics is along time-like geodesics. We further want to find an analogue of the massless field radiative phase space. It is then natural to seek for a smooth manifold like description of time-like infinity. A coordinate system that is well suited for this purpose is as follows.  Consider Minkowski space in spherical coordinates $(t,r,\xh)$ (we denote points in the sphere by unit vectors $\xh)$. For $t \geq r$ introduce the rescaled time and radial coordinates:
\ba
\t & := &\sqrt{t^2 - r^2}  ,\\
\rho & :=&  \frac{r}{\sqrt{t^2 - r^2}} ,
\ea
in terms of which the Minkowski metric reads:
\be
ds^2 = - d \t^2 + \t^2 \left( \frac{d \rho^2}{1+ \rho^2} + \rho^2 \gamma_{AB} d x^A d x^B \right), \label{metric}
\ee
where $A,B,\ldots$ denote sphere indices and $\gamma_{AB}$ is the unit sphere metric.  Eq. (\ref{metric}) takes the form $ds^2=-d \t^2 + \t^2 d\sigma^2$ with $d \sigma^2$ the metric of the unit space-like hyperbolid. We denote this (future) unit hyperboloid by $\H^+$. As we shall see  $\H^+$  represent the abstract manifold underlaying the asymptotic phase space of massive fields, in the same way as $\I^+$ underlays the asymptotic phase space of massless fields.

We now compute the  $\t \to \infty$  asymptotics of (\ref{freephi})  with $(\rho,\xh)$ fixed. Note that since $t = \t \sqrt{1+ \rho^2}$, this limit is equivalent to:
\be 
t \to \infty \quad {\rm with} \quad  \vec{v}:=\vec{x}/t \quad {\rm fixed}.
\ee
\\
In $(\tau,\rho,\xh)$ coordinates we have:
\be
 x \cdot p = \t(- \sqrt{1+ \rho^2} E_p + \rho \, \xh \cdot \vec{p}) =: \t f(\vec{p}) , \label{xdotp}
\ee
so that in the $\tau \to \infty$ limit the integral (\ref{freephi}) can be evaluated by a saddle point approximation. The relevant critical points of the function $f$ defined in (\ref{xdotp}) are:
\ba
\frac{\partial f}{\partial p_i}  = 0 \iff \vec{p}= m \rho \xh.
\ea
Using that $f(m \rho \xh) = -m$ and $\det(\partial_i \partial_j f(\vec{p}))=-m^2 (1+\rho^2)^{3/2} (E_p)^{-5} $ one finds:
\be
\varphi(\t,\rho,\xh) = \frac{e^{- i \pi/4}\sqrt{m}}{2 (2 \pi \tau)^{3/2}}  \left( b(m \rho,\xh) e^{ -i \t m } + i d^*(m \rho,\xh) e^{ i \t m } \right)+O(\t^{-5/2}) .  \label{asymphi}
\ee
We now describe the symplectic structure on the free data. Although in the current free field case we already know the answer (namely the one that corresponds to the  free field Poisson brackets), we rederive this result in a way that is applicable to the interacting field case discussed in the next section. The starting point is the covariant phase space symplectic product \cite{abr,lw}:
\be
\O_\Sigma(\delta,\delta')=\int_{\Sigma} dS_\mu \w^\mu(\delta,\delta'),
\ee
where
\be
\w^{\mu}(\delta,\delta') = - \sqrt{g} g^{\mu \nu}(\delta \varphi  \partial_\nu \delta' \varphi^* + c.c. - \delta \leftrightarrow \delta'),
\ee
and $\Sigma$ is any Cauchy surface.  We would like to evaluate this integral on the  $\tau=$ constant hyperboloids and  take $\tau \to \infty$. There is a subtlety here in that such hyperboloids are not  Cauchy surfaces: The $\tau=$ constant surfaces intersect null infinity at $u:=t-r=0$. In order to have a Cauchy surface we need to include the $u<0$ portion of $\I^+$ (we will shortly show that such portion does not contribute to the symplectic product). Let us denote by $\Sigma_\tau$ the resulting Cauchy surface, namely:
\be
\Sigma_\tau := \left\{
\begin{array}{cll}
\{t^2-r^2=\tau^2 \} & \text{for} & t \geq r  \\
\I^+  & \text{for}  & t< r .
\end{array} \right. 
\ee
For the $\I^+$ portion we need to consider the $(r \to \infty, u= {\rm constant})$ asymptotic form of the field $\varphi(x)$ (\ref{freephi}). A saddle point analysis shows that in this limit  $\varphi(x) =O(r^{-3/2})$ and $\w^r(\delta,\delta')=O(r^{-1})$. As the volume form on $\I^+$ is $dud^{2}z$ (where $z,\zb$ are complex coordinates on the conformal sphere)  we see that indeed the $\I^+$ contribution to the symplectic product vanishes.\footnote{ By contrast in the massless case one has $\varphi(x) =O(r^{-1})$ and $\w^r=O(1)$.} To evaluate the $\tau=$ constant hyperboloid contribution we use the asymptotic form (\ref{asymphi}) together with $\sqrt{g}= \t^3 \rho^2 (1+\rho^2)^{-1/2} \sqrt{\gamma}$ and $g^{\t\mu} = - \delta^{\t \mu}$. One finds:
\be
\w^\t(\delta,\delta') =\frac{i m^2 \sqrt{\gamma}}{2 (2 \pi)^3} \rho^2(1+\rho^2)^{-1/2}\left( \delta b \, \delta' b^*+  \delta d \, \delta' d^* - \delta \leftrightarrow \delta' \right) +O(\tau^{-1}). \label{wtauphi}
\ee
We thus conclude that:
\be
\lim_{\tau \to \infty} \Omega_{\Sigma_\tau}(\delta,\delta') = \frac{i m^2}{2 (2 \pi)^3} \int_{\H^+} d^3V \left( \delta b \, \delta' b^*+  \delta d \, \delta' d^* - \delta \leftrightarrow \delta' \right) , \label{Ophi}
\ee
where $d^3V$ is the volume element of the unit hyperboloid.  The complex fields $(b,d)$ viewed as functions on $\H^+$ with symplectic product given by the RHS of (\ref{Ophi}) define what we call the asymptotic phase space of the massive field. From now on this phase space will be denoted by $\Gamma^\phi$.

  By making  the change of variables $\vec{p}= \rho \xh$ one verifies that (\ref{Ophi}) becomes the symplectic product associated to the standard free field Poisson brackets:
\be
\{b(\vec{p}),b^*(\vec{p}') \}=\{d(\vec{p}),d^*(\vec{p}') \}  = -i  (2 \pi)^3 (2 E_p) \delta^3(\vec{p}-\vec{p}').
\ee

\section{Asymptotic phase space of scalar electrodynamics in Lorenz gauge}
We now consider the interacting system of a massive charged scalar $\varphi$ and Maxwell field $\A_\mu$.  For definitiveness we  work in Lorenz gauge $\nabla^\mu \A_\mu=0$. The field equations are:
\be
\nabla^\nu \F_{\mu \nu} =  \J_\mu , \quad  (-D_\mu D^\mu +m^2) \varphi=0, \label{eom}
\ee
where $\F_{\mu \nu}= \partial_\mu \A_\nu-\partial_\mu \A_\nu$, $D_\mu$  is the gauge covariant derivative,
\be
D_\mu \varphi = \partial_\mu \varphi - i e \A_\mu \varphi,
\ee
and $\J_\mu$ the current of the charged field,
\be
\J_\mu= i e \varphi (D_\mu \varphi)^* + c.c.
\ee
As usual in the asymptotic treatment of nonlinear theories, we need to impose  asymptotic conditions on the fields that are compatible with the field equations. Specifically,  we need conditions over the null $(t \to \infty, \, u=t-r= {\rm const.})$ and time-like $(t \to \infty, \, r/t= {\rm const}.)$ limits of the fields. At null infinity, we assume the fields obey the same leading fall-offs as free fields: 
\be
\A_u= r^{-1} A_u(u,\xh) + O(r^{-2}), \;  \A_r = O(r^{-2}), \; \A_A = A_A(u,\xh) + O(r^{-1}), \label{gralasymA}
\ee
\be
\varphi = O(r^{-3/2}), \quad \J_u = O(r^{-3}), \; \J_r=O(r^{-4}), \; \J_A=O(r^{-2}). \label{gralasymphi}
\ee
Conditions (\ref{gralasymA}) are obtained from a saddle point evaluation of the Fourier expansion of a free Maxwell field in Lorenz gauge (see for instance \cite{stromqed}). Conditions (\ref{gralasymphi}) can also be obtained from a saddle point analysis as mentioned in the previous section.  Substituting these expansions in the field equations (\ref{eom}) one finds that $A_A(u,\xh)$ is undetermined (thus representing free data) and  $A_u(u,\xh)$ satisfies:\footnote{A charged massless field would have given a contribution to the right hand side of \ref{Au}, as in  Eq. (2.9) of \cite{stromqed}.}
\be
\partial_u A_u = D^A \partial_u A_A. \label{Au}
\ee
We will further assume the `free data'  $A_A(u,\xh)$ satisfies the standard radiative phase space fall-offs:
\be
A_A(u,\xh)=A^{\pm}_A(\xh) +O(u^{-\epsilon}). \label{fallAu}
\ee

To describe  time-like asymptotics we switch to $(\rho,\tau)$ coordinates as defined in the previous section. Using (\ref{gralasymA}) and (\ref{gralasymphi}) one finds that in these coordinates the $\tau \to \infty, \rho= {\rm const.}$ asymptotics of the Maxwell field is:
\be \label{asymAtau}
\begin{array}{ccl}
\A_\tau & = &  A_\tau(\rho,\xh) \tau^{-1} + O(\tau^{-1-\epsilon})  \\
 \A_\rho & = & A_\rho(\rho,\xh) +O(\tau^{-\epsilon}),  \\
 \A_A & = & A_A^+(\xh)+O(\tau^{-\epsilon}),
 \end{array}
\ee
where $A_\tau$ and $A_\rho$ can be written in terms of $A_u(\infty,\xh)$ and certain $\rho$-dependent factors.  These fall-offs imply that the leading difference between the free and gauge covariant Klein-Gordon equation is:
\be
D^\mu D_\mu = \square - 2 i e A_\tau(\rho,\xh) \tau^{-1} \partial_\tau + \ldots .
\ee
The extra term has the effect of modifying the asymptotic form of the free massive field (\ref{asymphi}) to:
\be\label{asymptote}
\varphi  \propto \tau^{-3/2} e^{i \ln(\tau) A_\tau(\rho,\xh)} \big( b(\rho,\xh) e^{ -i \t m } + i d^*( \rho,\xh) e^{ i \t m }) + \ldots. 
\ee
As we shall see, this modification with respect to the asymptotic free   field (\ref{asymphi}) does not affect the expression of the symplectic product when expressed in terms of the `free data' $b(\rho,\xh)$ and $d(\rho,\xh)$.\footnote{We expect that the asymptotic behavior  (\ref{asymptote}) matches with the asymptotic fields used to define IR finite S-matrix in QED \cite{KF,akhoury}. Detailed investigation of this issue is left for the future.}\\

We now describe the symplectic structure of the Maxwell+massive field system. The covariant phase space symplectic density reads:
\be
\w^\mu(\delta,\delta')= \sqrt{g}(\delta \F^{\mu \nu} \delta' \A_\nu + (D^\mu \delta \varphi)^* \delta' \varphi + c.c.) - \delta \leftrightarrow \delta' \label{deftheta}.
\ee
Given a solution to the field equation $(\A_\mu,\varphi)$ and variations  $\delta,\delta'$ thereof, we want to evaluate the symplectic product in terms of the asymptotic fields by:
\be
\O(\delta,\delta'):= \lim_{t \to \infty} \int_{\Sigma_t} dS_\mu \w^\mu(\delta,\delta'),
\ee
with $\Sigma_t$ a $t=$ constant Minkowski time slice. The asymptotic form of $\w^t$ depends on how the $t=$ constant fields are parametrized in the radial direction as $t \to \infty$. If one keeps $u=t-r$ constant, conditions (\ref{gralasymA}) and (\ref{gralasymphi}) imply:
\begin{equation}
\begin{array}{lll}
\w^{t}\ =\ \w^{r}+\w^{u}\\
\vspace*{0.1in}
\w^{r}\ =\ \ \sqrt{\gamma} \gamma^{MN}\partial_{u} \delta' A_{N} \delta A_{M}\ - \delta \leftrightarrow \delta' +O(t^{-1})\\
\vspace*{0.1in}
\w^{u}\ =\ O(t^{-1}).
\end{array}
\end{equation}
If on the other hand one keeps $r/t$ constant, conditions (\ref{asymAtau}) and (\ref{asymptote}) imply that $\w^t$ coincides with the free massive field symplectic density (\ref{wtauphi}) up to terms that vanish in the $t \to \infty$ limit. We thus conclude that:
\be
\O(\delta,\delta') = \O_{A}(\delta,\delta') + \O_{\phi}(\delta,\delta'),
\ee
where
\be
\O_{A}(\delta,\delta') = \int_{\I^+}\sqrt{\gamma} du ( \delta A_A \partial_u  \delta' A^A - \delta \leftrightarrow \delta') 
\ee
is the standard  symplectic product of the Maxwell field radiative phase space $\Gamma^A$ and $\O_{\phi}(\delta,\delta')$  the free massive field symplectic product as given in the RHS of Eq. (\ref{Ophi}).

This provides the (future) asymptotic phase space of scalar electrodynamics. Similar considerations apply for the past asymptotic phase space.

\section{Asymptotic symmetries and associated charges}\label{asympsymm}
We thus see that the full phase space of the system which is relevant for defining the scattering states is given by $\Gamma\ :=\ \Gamma^{\phi}\times\Gamma^{A}$. In the Lorenz gauge we are considering, the theory has a residual  gauge invariance
\be \label{mar140}
\delta_{\lamt}{\cal A}_{\mu} = \partial_{\mu}\lamt  , \quad \delta_{\lamt}\varphi  = ie\lamt \, \varphi
\ee
with  gauge parameter satisfying   the wave equation
\begin{equation}\label{mar141}
\square\,  \lamt\ =\ 0.
\end{equation}
The asymptotic symmetries are by definition the large gauge transformations which are nontrivial at infinity. Let us for concreteness focus on future infinity. Preservation of fall-offs (\ref{gralasymA}) and (\ref{asymAtau}) impose the following null and time-like asymptotic conditions on $\lamt$:
\ba
\lamt(u,r,\xh) & = & \lambda(\xh) +O(r^{-1}), \\
\lamt(\tau,\rho,\xh) & = & \laminf(\rho,\xh) + O(\tau^{-\epsilon}) ,
\ea
for some functions $\lambda$ on the sphere and $\laminf$ on the hyperboloid.  Compatibility between the null and time-like limits requires that:
\be
\lim_{\rho \to \infty}  \laminf(\rho,\xh) = \lambda(\xh). \label{limlamH}
\ee
Since $\lamt$ satisfies the wave equation, $\lambda(\xh)$ provides the unconstrained data that determines $\lamt$. On the other hand, the  wave equation implies that $\laminf$ satisfies Laplace equation on $\H^+$:
\be
\Delta \laminf(\rho,\xh) =0, \label{deltalam}
\ee
where  $\Delta$ is the Laplacian on $\H$ (the wave operator in $(\rho,\tau)$ coordinates is: $\square = -\partial^2_\tau + \tau^{-2} \Delta$).

Conditions (\ref{limlamH}) and (\ref{deltalam}) determine $\laminf$ in terms of $\lambda$ according to:
\begin{equation}
\laminf(\rho,\xh) =\ \int_{S^{2}}d^{2} \qh \, {\cal G}(\rho,\xh;\qh)\ \lambda(\qh) ,
\end{equation}
where ${\cal G}$ is an integration kernel satisfying
\begin{equation} \label{defcalG}
\begin{array}{lll}
\Delta{\cal G}(\rho,\xh;\qh)\ =\ 0\\
\vspace*{0.1in}
\lim_{\rho\rightarrow\infty}{\cal G}\left(\rho,\xh; \qh \right)\ =\ \delta^{2}(\xh,\qh).
\end{array}
\end{equation}
To summarize: The large gauge transformations are parametrized by functions on the sphere $\lambda(\xh)$. 

From Eq. (\ref{mar140}) we obtain the following action of large gauge transformations on the `free data' asymptotic fields:
\ba
\delta_{\lambda}A_{A}(u,\xh) & =&  \partial_A \lambda(\xh) \\
\delta_{\lambda} b({\vec{p})} & = & i e  \laminf(\vec{p}/m) b(\vec{p})  \label{dellamb}\\
\delta_{\lambda} d({\vec{p})} & = & - i e  \laminf(\vec{p}/m) d(\vec{p}), \label{dellamd}
\ea
where $\vec{p}/m$ denotes the point $(\rho,\xh)=(|\vec{p}|/m, \vec{p}/|\vec{p}|)$ on $\H$. In the asymptotic phase space $\Gamma^{\phi}$, the generators of the large gauge transformations are given by:\\
\be
Q^\H_\lambda\ =- \int_{{\cal H}^{+}}d^{3}V\ \laminf(\rho,\xh)j^{\tau}(\rho,\xh) ,
\ee
where $j^\tau$ is the charge current across $\H^+$ defined by:
\be
j^\tau(\rho,\xh) = \lim_{\tau \to \infty} \tau^3 \J^\tau(\tau,\rho,\xh) = \frac{e m^2}{2 (2 \pi)^3}(b(\rho,\xh) b^*(\rho,\xh) -d(\rho,\xh) d^*(\rho,\xh)).
\ee

On the other hand, 
the contribution to the charges from null infinity
were derived in 
\cite{AS,stromqed}
and are given by
\begin{equation}
Q^\I_\lambda =
\int_{\I^{+}}du \sqrt{\gamma} \lambda(\xh)\left[\partial_{u}(D^A A_A)-\ j_{u}\right],
\end{equation}
where $j_{u}\ =\lim_{r\rightarrow\infty}(r^{2}\J_{u})$ is the component of the charge current along $\I^{+}$. As there are no massless charged particles in the theory, this current vanishes and whence the total charge which generates large gauge transformations in the asymptotic phase space $\Gamma$ is: 
\ba
Q_\lambda & = &  -\int_{{\cal H}^{+}}d^{3}V\ \lambda_\H \, j^{\tau} + \int_{S^2}\sqrt{\gamma} \,  \lambda D^A[ A_A] \\
& =: & \Qh_\lambda+\Qs_\lambda
\ea
where  we wrote the Maxwell field contribution  as a $u = \pm \infty$ boundary term and
\be 
[A_A](\xh) := A^+_A(\xh) - A^-_A(\xh).
\ee
Following the standard terminology, the contribution to $Q_\lambda$ that is quadratic in the fields is referred to as `hard'  and the one linear in the gauge fields as `soft'. \\ 
 
There are two final  ingredients that are needed to prove  the equivalence between Ward identities and soft theorem (these ingredients are also present in the supetranslation case of gravity). The first one is a requirement on the Maxwell radiative data, namely:
\be
[F_{z \zb}] =0 \iff D^z[A_z]=D^{\zb}[A_{\zb}]  .\label{cond}
\ee
 We will see that although this condition is implied by the soft theorem, it is an hypothesis needed when deducing the soft theorem from the Ward identity.  Under condition (\ref{cond}), the soft charge can be written as:
\ba
\Qs_\lambda & = & - 2 \int d^2 z \sqrt{\gamma}  \lambda D_z [A^z] ,\label{Qsz}\\
&=&  - 2 \int d^2 z \sqrt{\gamma}  \lambda D_{\zb}[A^{\zb}]. \label{Qszb}
\ea
The second ingredient is a rewriting of $[A_A(\xh)]$ (and hence of the soft charge) in terms of the Fourier transform of $A_A(u,\xh)$,
\be
A_{A}(E,\xh) = \int_{-\infty}^\infty  A_{A}(u,\xh) e^{i E u} du ,
\ee
according to:\footnote{From the classical radiative phase space perspective, there is no need to specify how the limit $E_s \to 0$ is taken. In quantum theory however, the prescription provides the precise definition of the soft charge. We choose (\ref{prescription}) for convenience, but in fact any prescription of the form $[A_A(z,\zb)] =-i  \lim_{E_s \to 0^+} E_s (a  A_A(E_s,z,\zb)- (1-a) A_A(-E_s,z,\zb))$ can be used to show the equivalence. Perhaps physically the most natural choice is $a=1/2$ as used in \cite{stromqed}.}
\be
[A_A(\xh)] = -i \lim_{E_s \to 0^+}  E_s A_A(E_s,\xh). \label{prescription}
\ee
\\

This concludes the discussion of asymptotic symmetries and charges at future infinity. Similar discussions goes through for past infinity. Since $\lamt$ satisfies the free wave equation (\ref{mar141}), its value at $\I^-$ coincides with that of $\I^+$ by an antipodal identification of the future and past conformal spheres. Thus,  the total (future and past) group of asymptotic symmetries coincides with the corresponding group  of massless QED \cite{stromqed}, and is parametrized by a single sphere function $\lambda(\xh)$.

\section{Equivalence between Ward Identities and soft theorem}
Our starting point is the formal\footnote{This S-matrix is well known to be IR divergent.} S-matrix of QED, which is defined by the asymptotic (free) in and out states of the massive scalar and Maxwell fields. We do believe that the proper context to understand asymptotic symmetries of QED should be the IR finite S-matrix associated to  Kulish-Faddeev coherent states \cite{KF}. We refer to the work \cite{balsachin} which analyzes  asymptotic symmetries in the context of such coherent states. In the radiative phase space framework 
 the coherent states were described  in \cite{aa,narain}. We believe there is a close connection between these works and the ideas in this paper but investigating these connections is outside the scope of the present paper.\\
Whence in this section we start by exploring the consequence of the conjecture that symmetry group of QED S-matrix is the group of large $U(1)$ gauge transformations.
A given QED vacuum state corresponds to spontaneous breaking of this group to global $U(1)$ and the soft photons are the associated goldstone bosons. We write down the Ward identities corresponding to this spontaneously broken symmetry and show that these identities are \emph{equivalent} to Weinberg's soft photon theorem.

\subsection{Ward Identity}
The starting point is the conjectured identity \cite{stromqed}: $Q_\lambda^+ S = S Q_\lambda^-$ which one rewrites as:
\be
Q^{\rm soft +}_\lambda S- S Q^{\rm soft -}_\lambda = - Q^{\rm hard +}_\lambda S + S  Q^{\rm hard -}_\lambda. \label{wi0}
\ee
One then evaluates the matrix elements of (\ref{wi0}) between asymptotic states $|{\rm in} \ket$ and  $\bra {\rm out}|$  composed of charged particles. In the interaction picture such states belong to the Fock space associated to the free field operators $b(\vec{p})$ and $d(\vec{p})$. The hard charge is then defined by a normal ordered version of the classical expression. Its action on the asymptotic states is determined by the commutations relations:
\be
[b(\vec{p}),\Qh_\lambda]= - e \laminf(\vec{p}/m) b(\vec{p}), \quad [d(\vec{p}),\Qh_\lambda]= e \laminf(\vec{p}/m) d(\vec{p}) \label{bQ},
\ee
which represent the quantum version of relations (\ref{dellamb}), (\ref{dellamd}). For the Maxwell field, the standard  free field Fock operators $a_\pm(E,z,\zb)$ are related to $A_A(E,z,\zb)$ ($E>0$) by \cite{stromqed}:
\be
A_z(E,z,\zb) = \frac{\sqrt{\gamma_{z\zb}}}{4 \pi i} a_+(E,z,\zb) , \quad A_{\zb}(E,z,\zb) = \frac{\sqrt{\gamma_{z\zb}}}{4 \pi i} a_-(E,z,\zb), \label{Aw}
\ee
where $\gamma_{z \zb}=2/(1+z \zb)^2$. Using (\ref{Qsz}), (\ref{prescription}) and (\ref{Aw}) we write the soft charge operator as:
\be
\Qs_\lambda= \lim_{E_s \to 0^+} \frac{E_s}{2 \pi} \int d^2 w \lambda(w,\wb) \partial_{\wb} (\sqrt{\gamma_{w \wb}} a_+(E_s,w,\wb)) . \label{softQ}
\ee
From (\ref{softQ}) and the commutation relations (\ref{bQ}) the matrix element of (\ref{wi0}) is found to be:
\be
 \lim_{E_s \to 0^+} \frac{E_s}{2 \pi} \int d^2 w  \lambda(w,\wb) \partial_{\wb} \big(\sqrt{\gamma_{w \wb}}\bra {\rm out} |a_+(E_s,w,\wb) S | {\rm in} \ket \big) = e \sum_i q_i \laminf(\vec{p}_i/m)\bra {\rm out} |  S | {\rm in} \ket ,  \label{wi}
\ee
where the sum runs over all external particles, with $q_i=+1$  for outgoing particles/incoming antiparticles  and  $q_i=-1$ for incoming particles/outgoing antiparticles. Below we show that (\ref{wi}) is  equivalent to the soft theorem.
\subsection{From soft theorem to Ward identity}
In the notation of the previous subsection, Weinberg's soft photon theorem for an outgoing positive helicity soft photon reads:
\be
\lim_{E_s \to 0^+} E_s  \bra {\rm out} | a^{\rm out}_{+} (E_s,w,\wb) S | {\rm in} \ket  =e \sum_{i}q_i \frac{\e^+(w,\wb) \cdot p_i}{(q/E_s) \cdot p_i} \bra {\rm out} |  S | {\rm in} \ket , \label{softthm}
\ee
where  $q/E_s \equiv (1,\qh)$ with $\qh$ parametrized by $(w,\wb)$ and $\e^+(w,\wb)$ the polarization vector \cite{stromqed}:
\be
\e^{+ \mu}(w,\wb)=1/\sqrt{2}(\wb,1,-i,-\wb).
\ee
We now perform the operation $(2 \pi)^{-1} \int d^2 w  \lambda(w,\wb) \partial_{\wb} \big(\sqrt{\gamma_{w \wb}} \, $ on both sides of (\ref{softthm}). The LHS becomes the left term in (\ref{wi}). The RHS takes the same form as the right term in (\ref{wi}), with a multiplier $\lambda'(\vec{p}/m)$ given by:
\be
\lambda'(\vec{p}/m) = \int d^2 w \, G(\vec{p}/m; w,\wb)  \, \lambda(w,\wb), \label{lambdap}
\ee
\be
 G(\vec{p}/m; w,\wb) := \frac{1}{2\pi}\partial_{\wb} \left(\sqrt{\gamma_{w \wb}}\frac{\e^+(w,\wb) \cdot p}{(q/E_s) \cdot p}\right). \label{partialsoft}
\ee
Parametrizing the 3-momentum particle as  $\vec{p}= m \rho \xh$ one can verify that (\ref{partialsoft}) takes the following simple form:
\be
G(\rho,\xh ; \qh) = \frac{\gamma^{1/2}(w,\wb)}{4 \pi} (\sqrt{1+\rho^2}-\rho \qh \cdot \xh)^{-2}.\label{Gdwb}
\ee
We now show that  $G(\rho,\xh ; \qh)$ is nothing but the  kernel $\G$  discussed in section \ref{asympsymm}.   First, by direct computation one can verify that $G$ satisfies the Laplace equation on the hyperboloid $(\rho,\xh)$:
\be
\Delta G(\rho,\xh ; \qh) =0 \label{lapzero},
\ee
where the explicit coordinate expression of the Laplacian is:
\be
\Delta= (1+\rho^2) \partial^2_\rho + \rho^{-1}(2+3 \rho^2) \partial_\rho +\rho^{-2} (1+ z \zb)^2 \partial_z \partial_{\zb} 
\ee
($(z,\zb)$ parametrizes $\xh$).  Second, we note that:\footnote{This limit is equivalent to the $m \to 0, \, \vec{p}= $ constant limit, and thus the expression coincides with the soft factor for  massless particles.}
\be
\lim_{\rho \to \infty}\sqrt{\gamma_{w \wb}}\frac{\e^+(w,\wb) \cdot p}{(q/E_s) \cdot p} =(w-z)^{-1} .\label{limrho}
\ee
 From (\ref{partialsoft}), (\ref{limrho}) and  $\partial_{\wb}(w-z)^{-1}= 2 \pi \delta^{2}(w-z)$ we conclude that:
\be
\lim_{\rho \to \infty} G(\rho, z,\zb ; w,\wb) =  \delta^{2}(w-z).
\ee
It then follows that $G$ satisfies the defining relations (\ref{defcalG}) for $\G$ and hence $\lambda'(\rho,\xh)$ coincides with $\lambda_\H(\rho,\xh)$ as defined in section \ref{asympsymm}.  This completes the derivation of the Ward identity (\ref{wi}) from the soft theorem (\ref{softthm}).


We conclude by noting that the fact  that $G(\rho,\xh ; \qh)$ is real implies condition (\ref{cond}). Starting with  the soft theorem for a negative helicity soft photon and repeating the same operation as before (with $\partial_w$ in place of $\partial_{\wb}$), one obtains an equation as in (\ref{wi}) with $\lambda'=\int d^2 w \, G^*(\vec{p}/m; w,\wb)  \, \lambda(w,\wb)$ (here  we used the fact that the negative helicity polarization vector is the complex conjugated of the positive helicity one). Since $G$ is real, we obtain the same $\lambda'$ as before. As this holds true for any choice of $\lambda$ on the sphere, one concludes that:
\be
\lim_{E_s \to 0^+} E_s \partial_{\wb} (\sqrt{\gamma_{w \wb}}a_+(E_s,w,\wb) ) = \lim_{E_s \to 0^+} E_s  \partial_w (\sqrt{\gamma_{w \wb}} a_-(E_s,w,\wb)), \label{cond2}
\ee
which corresponds to the classical condition (\ref{cond}).

\subsection{From Ward identity to soft theorem}
Consider the Ward identity (\ref{wi}) with
\be
\lambda(w,\wb) = (z_s-w)^{-1}. \label{lamwi}
\ee
By integration by parts, the LHS of the Ward identity becomes the LHS of the soft theorem (\ref{softthm}) (times an overall multiplicative factor $\sqrt{\gamma_{z_s \zb_s}}$) with  soft photon direction given by $z_s$. The gauge parameter on $\H$ determined by (\ref{lamwi}) is:
\ba
\lambda_\H(\vec{p}/m) &= &  \int d^2 w \, G(\vec{p}/m; w,\wb)  \,  (z_s-w)^{-1}  , \\
&=&  \sqrt{\gamma_{z_s \zb_s}}\frac{\e^+(z_s,\zb_s) \cdot p}{(q/E_s) \cdot p} ,
\ea
where we used (\ref{partialsoft}) and made an integration by parts. As before $q/E_s\equiv(1,\qh(z_s,\zb_s))$. We thus obtain the RHS of (\ref{softthm}) (times the overall multiplicative factor $\sqrt{\gamma_{z_s \zb_s}}$) and recover the soft theorem for a positive helicity photon. The soft theorem for a negative helicity photon is similarly obtained by starting with the version of the Ward identity that uses the form (\ref{Qszb}) of the soft charge, and by taking $\lambda(w,\wb) = (\zb_s-\wb)^{-1}$.
\section{Conclusions and Outlook}
In this paper we have explored the consequences of an infinite dimensional group of large gauge transformations being a symmetry group of massive QED. We showed that precisely as in the case of massless QED \cite{stromqed}, the Ward identities associated to these asymptotic symmetries are equivalent to Weinberg's soft photon theorem \cite{weinberg}. 

There are two  key ingredients that allowed us to extend the analysis of massless charges given in \cite{stromqed}  to the case of  massive charges. 
 The first one is a suitable definition of time-like infinity that  provides a kinematical arena for the  asymptotic phase space of massive fields.  The situation here is  analogous to how the radiative phase space of massless fields is defined at null infinity \cite{AS}.

  The second ingredient is the use of a gauge fixing condition that is well behaved at both null and time-like infinity. Here Lorenz gauge proved to be a convenient choice which furthermore facilitated the comparison with Weinberg's soft factors. 
On this issue it is  important to  emphasize that the soft theorem, the Ward identities, and their equivalence  are all gauge invariant statements.  The gauge choice features in the specific form taken by individual terms of an otherwise gauge invariant sum.  
On the soft theorem side, the gauge choice is reflected in the polarization vectors of the photons. On the Ward identity side, the gauge choice determines  how the sphere function $\lambda(\xh)$ extends  inside the unit hyperboloid that describes  time-like infinity.  
\\

One of the main motivations behind recent resurgence in the field of asymptotic symmetries in QFTs is to understand the symmetry group of quantum gravity in asymptotically flat spacetimes. It is well known since 60s that the symmetry group of asymptotically flat solutions to Einstein's equations is the  BMS group which is a semi-direct product of so called supertranslations (``angle dependent" translations along null infinity) and the Lorentz group. By postulating the BMS group (in fact a diagonal subgroup of the two BMS groups which are associated to future and past null infinities) Strominger and collaborators showed that the Ward identities associated to the infinite dimensional Abelian group of supertranslations is equivalent to the soft graviton theorem. Their analysis was done for the case where the gravitating particles are massless and hence all the scattering states were defined at null infinity. It is therefore a natural question to ask if, by postulating BMS group as a symmetry of (perturbative) quantum gravity coupled to massive fields, can one derive similar remarkable consequences like the soft graviton theorems. Based on the analysis done in this paper, and using the asymptotic phase space of massive scalar field on which 
the action of (late time) diffeomorphisms associated to supertranslations could be defined, we find this to be highly plausible and hope to come back to it in near future.

\section{Acknowledgements}
We are grateful to A.P. Balachandran and Sachin Vaidya for detailed discussions on the group of large gauge transformations in QED. MC is supported by Anii and Pedeciba. AL is supported by Ramanujan Fellowship of the Department of Science and Technology.

\end{document}